\documentclass[prb,preprint,groupedaddress]{revtex4}
\usepackage{graphicx}
\begin{document}
\bibliographystyle{prbrev}

\title{The structure and energetics of $^3$He and $^4$He nanodroplets doped
with alkaline earth atoms}

\author{Alberto Hernando}
\affiliation{Departament ECM, Facultat de F\'{\i}sica,
and IN$^2$UB,
Universitat de Barcelona. Diagonal 647,
08028 Barcelona, Spain}

\author{Ricardo Mayol}
\affiliation{Departament ECM, Facultat de F\'{\i}sica,
and IN$^2$UB,
Universitat de Barcelona. Diagonal 647,
08028 Barcelona, Spain}

\author{Mart\'{\i} Pi}
\affiliation{Departament ECM, Facultat de F\'{\i}sica,
and IN$^2$UB,
Universitat de Barcelona. Diagonal 647,
08028 Barcelona, Spain}

\author{Manuel Barranco}
\affiliation{Departament ECM, Facultat de F\'{\i}sica,
and IN$^2$UB,
Universitat de Barcelona. Diagonal 647,
08028 Barcelona, Spain}

\author{Francesco Ancilotto}
\affiliation{
INFM-DEMOCRITOS
and Dipartimento di Fisica  `G. Galilei', Universit\`a di Padova,
via Marzolo 8, I-35131 Padova, Italy
}

\author{Oliver B{\"u}nermann}
\affiliation{Physikalisches Institut, Universit\"at Freiburg.
Hermann-Herder-Str. 3, D-76104 Freiburg, Germany}

\author{Frank Stienkemeier}
\affiliation{Physikalisches Institut, Universit\"at Freiburg.
Hermann-Herder-Str. 3, D-76104 Freiburg, Germany}

\begin{abstract}
We present systematic results, based on density functional calculations, for the structure and energetics of 
$^3$He and $^4$He nanodroplets doped with alkaline earth atoms. We predict that alkaline earth atoms from Mg to 
Ba go to the center of $^3$He drops, whereas Ca, Sr, and Ba reside in a deep dimple at the surface of $^4$He drops, 
and Mg is at their center. For Ca and Sr, the structure of the dimples is shown to be very sensitive to the 
He-alkaline earth pair potentials used in the calculations. The $5s5p\leftarrow5s^2$ transition of strontium atoms 
attached to helium nanodroplets of either isotope has been probed in absorption experiments. The spectra 
show that strontium is solvated inside $^3$He nanodroplets, supporting the calculations. In the light of our findings, 
we emphasize the relevance of the heavier alkaline earth atoms for analyzing mixed $^3$He-$^4$He nanodroplets, 
and in particular, we suggest their use to experimentally probe the $^3$He-$^4$He interface.\\

\end{abstract}

\maketitle

\section{Introduction}

Optical investigations of impurities in liquid helium have
drawn considerable attention in the past.\cite{Tab97} In recent years,
experiments involving helium nanodroplets have added new
input into the interaction of atomic impurities with a superfluid
helium environment.\cite{Sti01,Sti06} In particular, the shifts of the electronic
transition lines represent a very useful observable to determine
the location of the foreign atom attached to a helium drop.

While most impurities are found to reside in the interior of
helium droplets,\cite{Bar93,Dal94,Cal01,Toe04} it is well-established that alkali atoms, due
to their weak interaction with helium, reside in a `dimple' at
the surface of the drop for both helium isotopes.\cite{Anc95,Sti04,May05} The
question of solvation versus surface location for an impurity
atom in liquid He can be addressed within the model of Ref.~\onlinecite{Anc95a},
where a simple criterion has been proposed to decide whether
surface or solvated states are energetically favored. An adimensional
parameter $\lambda$ can be defined in terms of the impurity-
He potential well depth $\epsilon$ and the minimum position $r_{min}$,
namely, $\lambda \equiv \rho \,\epsilon \, r_{min}/(2^{1/6}\sigma)$, where $\rho$ and $\sigma$ are the density and
surface tension of bulk liquid He, respectively. The threshold
for solvation in 4He is\cite{Anc95a} $\lambda \sim \lambda _0$,
with $\lambda _0=1.9$. When $\lambda<\lambda_0$,
a stable state of the impurity on the droplet surface is expected,
whereas when $\lambda>\lambda _0$, the impurity is likely to be solvated in
the interior of the droplet. Impurities such as neutral alkali atoms,
that weakly interact with helium, are characterized by values
of $\lambda$ much smaller than the above threshold; their stable state is
thus expected to be on the surface of the droplet, as experimentally
found.

The shape of the impurity-He interaction potential, however,
is not given consideration by this model. For cases in which
the value of $\lambda$ does not lie near (say, within 0.5) the solvation
threshold $\lambda_0$, the shape of the potential surface does not need
to be taken into account, as the model is predictive outside of
this threshold window. However, for values which lie close to
$\lambda_0$, consideration of the shape of the potential energy surface,
as well as the well depth and equilibrium internuclear distance,
is mandatory, and more detailed calculations are needed to
ascertain whether the impurity is solvated or not. It is worth
noticing that the above criterion works for either helium isotope,
although so far, it has been applied to $^4$He because experimental
data for $^3$He only appeared recently.\cite{Gre98,Har01,Sti04,Bue06}

Among simple atomic impurities, alkaline earth (Ake) atoms
play a unique role. While, for example, all alkali atoms reside
on the surface and all noble gas atoms reside in the interior of
drops made of either isotope,\cite{Bar06} the absorption spectra of heavy
alkaline earth atoms Ca, Sr, and Ba attached to a $^4$He cluster
clearly support an outside location of Ca and Sr\cite{Sti97} and likely
also of Ba,\cite{Sti99} whereas for the lighter Mg atom, the experimental
evidence shows that it resides in the interior of the $^4$He
droplets.\cite{Reh00,Prz07}

According to the magnitude of the observed shifts, the dimple
in the case of alkaline earth atoms is thought to be more
pronounced than in the case of alkalis, indicating that alkaline
earth atoms reside deeper inside the drop than alkali atoms. This
will be corroborated by density functional calculations presented
in the Theoretical Results. Laser-induced fluorescence results
for Ca atoms in liquid $^3$He and $^4$He have been recently reported\cite{Mor05}
and have been analyzed using a vibrating `bubble model' and
fairly old Ca-He pair potentials based on pseudopotential SCF/
CI calculations.\cite{Czu91}

Applying the simple criterion described above, Ca and Sr appear to be barely stable
in their surface location with respect to the bulk one,\cite{Anc03} as reflected in the $\lambda$ values collected in Table
\ref{table1}, which are close to $\lambda_0$ for these doped $^4$He systems. This
borderline character for the solvation properties of these
impurities implies that detailed calculations are required to help
to understand the results of spectroscopic studies on alkalineearth-
doped He droplets. In particular, high-quality impurity-
He pair interaction potentials are required since even relatively
small inaccuracies in these potentials, which are often not known
with a sufficient precision, may yield wrong results.

We present here a systematic study for helium drops made
of each isotope, having a number of atoms large enough to make
them useful for the discussion of experiments on laser-induced
fluorescence (LIF) or beam depletion (BD) spectroscopy or for
the discussion of other physical phenomena involving these
systems, such as interatomic Coulombic decay.\cite{Ced97,Kry06} We also
discuss the dependence of the structural properties on the cluster
size. Some of this information is also experimentally available.\cite{Sti99}
After a brief explanation of the experiment, results are presented
for strontium on helium nanodroplets that further support the
calculations.

This work is organized as follows. In Sec.~\ref{ii} we briefly describe the density functional
plus alkaline earth-He potential approach employed here, as well as some technical
details. Doped drops calculations are presented and discussed in Sec.~\ref{iii}, while the
experimental results are discussed in Sec.~\ref{iv}, and an outlook is presented in Sec.~\ref{v}.

\section{Density-functional description of helium nanodroplets\label{ii}}

Since the pioneering work of Stringari and coworkers,\cite{Str87} density functional (DF)
theory has been used in many studies on liquid helium in confined geometries and found to
provide a quite accurate description of the properties of inhomogeneous liquid He (see
e.g. Ref.~\onlinecite{Bar06} and references therein).

The starting point is to write the  energy of the system as a functional of the He
particle density $\rho$:
\begin{equation}
E[\rho] = \int d{\bf r}\, {\cal E}(\rho)
+ \int d{\bf r}'\, \rho({\bf r}'\,) V_{Ake-He}(|{\bf r}-{\bf r}'|)  \; ,
\label{eq2}
\end{equation}
where ${\cal E}(\rho)$ is the He energy density per unit volume, and $V_{Ake-He}$ is the
alkaline earth-helium pair potential.
The impurity is thus treated as a fixed external
potential. Addressing the lightest alkaline earth, Be, for which  a fairly recent Be-He
is available,\cite{Par01} would likely require to treat this atom as a quantum particle
instead of as an external potential.\cite{Bar06}

For $^4$He we have used the Orsay-Trento functional,\cite{Dal95} and for $^3$He the one
described in Ref.~\onlinecite{May01} and references therein. These functionals have been
used in our previous work on helium drops doped with alkali atoms\cite{Sti04,May05} as
well as in many other theoretical works. The results discussed in the following have been
mostly obtained using the potentials of Ref.~\onlinecite{Lov04} (Ca, Sr and Ba), and  of
Ref.~\onlinecite{Hin03} (Mg, for which the pair potentials of Refs. \onlinecite{Par01}
and \onlinecite{Hin03} are similar). For Ca, we have also tested other potentials
available in the literature, \cite{Par01,Hin03,Czu03} as well as the unpublished
potential of Meyer\cite{Meyer} we had employed in our previous work.\cite{Anc03}

\begin{figure}[tb]
\centerline{\includegraphics[width=0.5\textwidth]{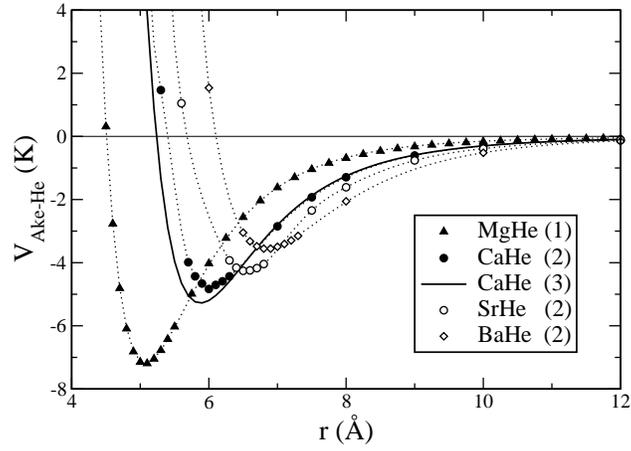}}
\caption{
Alkaline earth-He pair potentials used in this work to obtain the
ground state structure of doped helium drops:
(1)  Ref.~\onlinecite{Hin03};
(2)  Ref.~\onlinecite{Lov04};
(3)  Ref.~\onlinecite{Meyer}.
}
\label{fig1}
\end{figure}

Fig. \ref{fig1} shows the pair potentials used in this work. From this figure, one may
anticipate that Ca@$^4$He$_N$ drops described using the potential of Ref.
\onlinecite{Meyer} display deeper dimples than the same drops described with the
potential of  Ref.~\onlinecite{Lov04}. We want to point out that the Ca-He potentials of
Refs. \onlinecite{Par01} and \onlinecite{Hin03} are very similar to that of Ref.
\onlinecite{Lov04}, and should yield equivalent results. Contrarily, we have found that the
potential of Ref.~\onlinecite{Czu03} is more attractive, 
causing the Ca atom to be drawn to the center of the
$^4$He$_N$ drop, in contrast with the experimental findings.\cite{Sti99}

For a number $N$ of helium atoms in the drop, we have solved the Euler-Lagrange equation
which results from the variation of $E[\rho]$ at constant $N$:
\begin{equation}
\frac{\delta {\cal E}}{\delta \rho} + V_{Ake-He} = \mu  \; ,
\label{eq3}
\end{equation}
where $\mu$ is the helium chemical potential, whose value is determined self-consistently
by imposing the auxiliary condition $\int d{\bf r} \rho ({\bf r})=N$ during the iterative
minimization.

When the impurity resides off center (as in the case of a
dimple structure), the system is axially symmetric. Despite this
symmetry, we have solved Eq. (\ref{eq3})

in Cartesian coordinates because
this allows us to use fast Fourier transform techniques\cite{FFT} to
efficiently compute the convolution integrals entering the
definition of ${\cal
E}(\rho)$, that is, the mean field helium potential and
the coarse-grained density needed to evaluate the correlation
term in the density functional.\cite{Dal95} We have found this procedure
to be faster and more accurate than convoluting by direct
integration using cylindrical coordinates.

We have used an imaginary time method\cite{Pre92,Anc03a} to solve Eq. (\ref{eq3}), after
having discretized it using $13$-point formulas for the spatial derivatives.
The mesh used to discretize $\rho$ in space is chosen so that the results are
stable against small changes of the mesh step.

\section{Theoretical Results\label{iii}}

We start a typical calculation by placing the impurity close
to the surface of the He droplet. Depending on the studied
impurity and/or the He isotope, during the functional minimization,
the alkaline earth atom is either driven to the interior of the
droplet,\cite{note} or it remains trapped in a more or less pronounced dimple on its
surface.

\begin{figure}[tb]
\centerline{\includegraphics[width=0.5\textwidth]{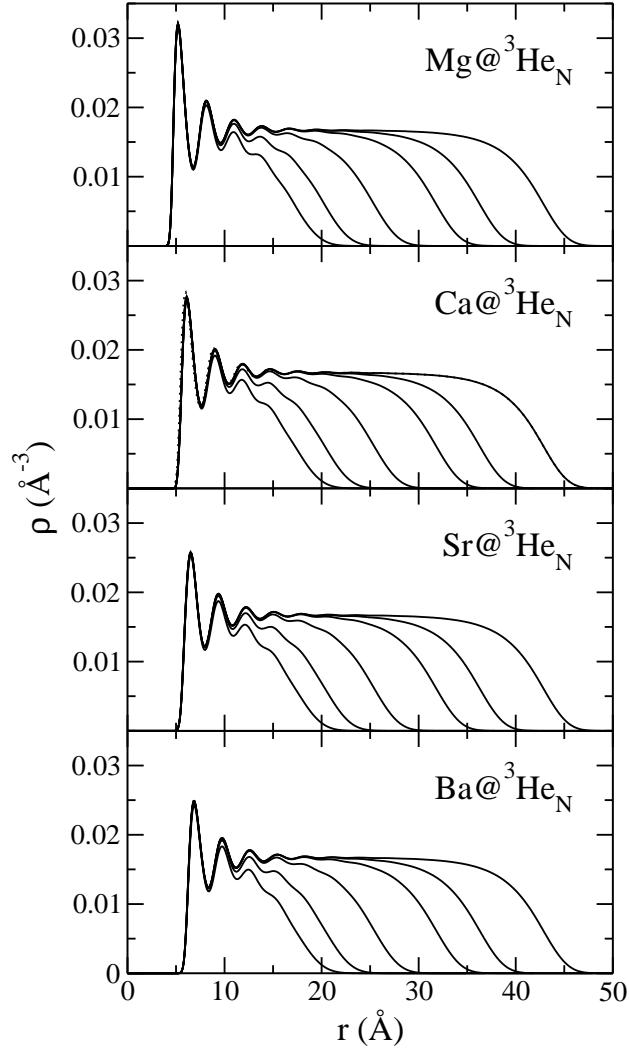}}
\caption{
Density profiles for $^3$He$_N$ drops doped with Mg, Ca, Sr, and Ba,
for $N=$ 300, 500, 1000, 2000, 3000, and 5000.
The dotted line in the Ca panel corresponds to Ca@$^3$He$_{5000}$
calculated with the pair potential of Ref.~\onlinecite{Meyer}.
Drops doped with
Ca, Sr and Ba have been calculated using the pair potentials of
Ref.~\onlinecite{Lov04}, and drops doped with Mg, using the pair potential of
Ref.~\onlinecite{Hin03}.
}
\label{fig2}
\end{figure}

In the case of $^3$He, we find that for all of the alkaline earth
atoms investigated, the stable state is always the one where the
impurity is in the center of the cluster. This is consistent with
the associated large $\lambda$ values, see Table
\ref{table1}.
Figure \ref{fig2} shows the density profiles for Mg@$^3$He$_N$, Ca@$^3$He$_N$,
Sr@$^3$He$_N$, and Ba@$^3$He$_N$  for $N=$ 300, 500, 1000, 2000, 3000, and 5000. For
Ca@$^3$He$_{5000}$, we also show the profile obtained with the pair potential of Ref.
\onlinecite{Meyer} (dotted line). Several solvation shells are clearly visible. The
number of $^3$He atoms below the first solvation peak for the $N=5000$ drop
is about 19 for
Mg, 22 for Ca, 26 for Sr, and 27 for Ba. The differences in the location
and height of the first solvation peak are a simple consequence
of the different depth and equilibrium distance of the corresponding
pair potentials. It is interesting to see the building up
of the drop structure around the impurity that, as it is wellknown,\cite{Dal94} only
causes a large but localized effect on the drop structure.

\begin{figure}[tb]
\centerline{\includegraphics[width=0.5\textwidth]{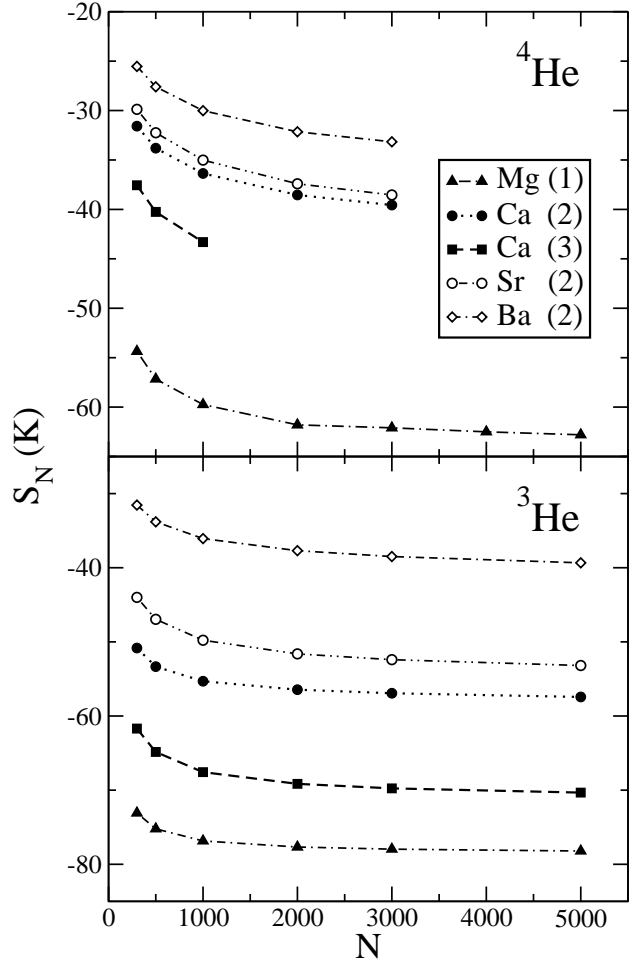}}
\caption{ Top panel: solvation energies (K) for doped $^4$He$_N$ drops.
Results obtained using the following pair potentials:
(1) from  Ref.~\onlinecite{Hin03};
(2) from  Ref.~\onlinecite{Lov04};
(3) from  Ref.~\onlinecite{Meyer}.
Bottom panel: same as top panel for doped $^3$He$_N$ drops.
The lines are drawn to guide the eye.
}
\label{fig3}
\end{figure}

The bottom panel of Fig. \ref{fig3} shows the corresponding solvation energies,
defined as the energy differences
\begin{equation}
S_N(Ake)= E(Ake@^3He_N) - E(^3He_N) \; ,
\label{eq4}
\end{equation}
with an equivalent definition for $^4$He drops. The more attractive Ca-He pair potential
of Ref.~\onlinecite{Meyer} yields, on average, a solvation energy about 13 K larger as
compared with that obtained with the pair potential of Ref.~\onlinecite{Lov04}, despite the fact that the density
profiles look fairly similar; see Fig. \ref{fig2}.

In the case of Ca and Sr atoms in $^4$He drops, whose $\lambda$ values are close to the
threshold for solvation $\lambda _0$ (see Table \ref{table1}), we
have found that, for both dopants, the minimum energy
configuration is a dimple state at the surface, although the energy difference between the surface and the solvated states
is fairly small for both dopants. For Ca@$^4$He$_{300}$, this difference is 3.4 K using
Meyer's potential,\cite{Anc03} and 12.0 K using that of Ref.~\onlinecite{Lov04}.
The homologous result for Sr@$^4$He$_{300}$ is  22.7 K. These energy differences
have to be compared with the total energy of the $^4$He$_{300}$ drop, which is about
$-$1384 K.

\begin{figure}[tb]
\centerline{\includegraphics[width=0.5\textwidth]{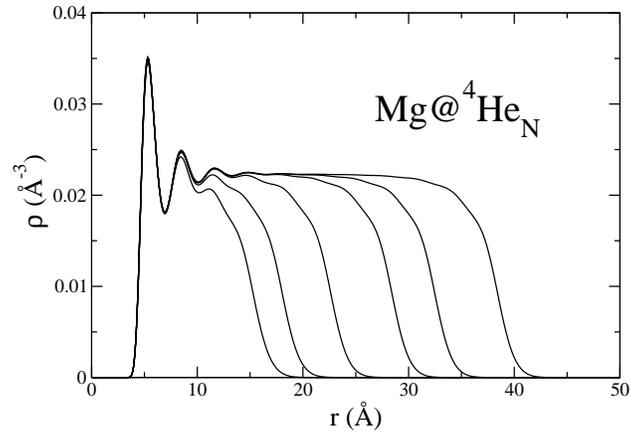}}
\caption{
Density profiles for Mg@$^4$He$_N$ drops
for $N=$ 300, 500, 1000, 2000, 3000, and 5000.
Results obtained using the pair potential of Ref.~\onlinecite{Hin03}.
}
\label{fig4}
\end{figure}

\begin{figure}[tb]
\centerline{\includegraphics[width=0.5\textwidth]{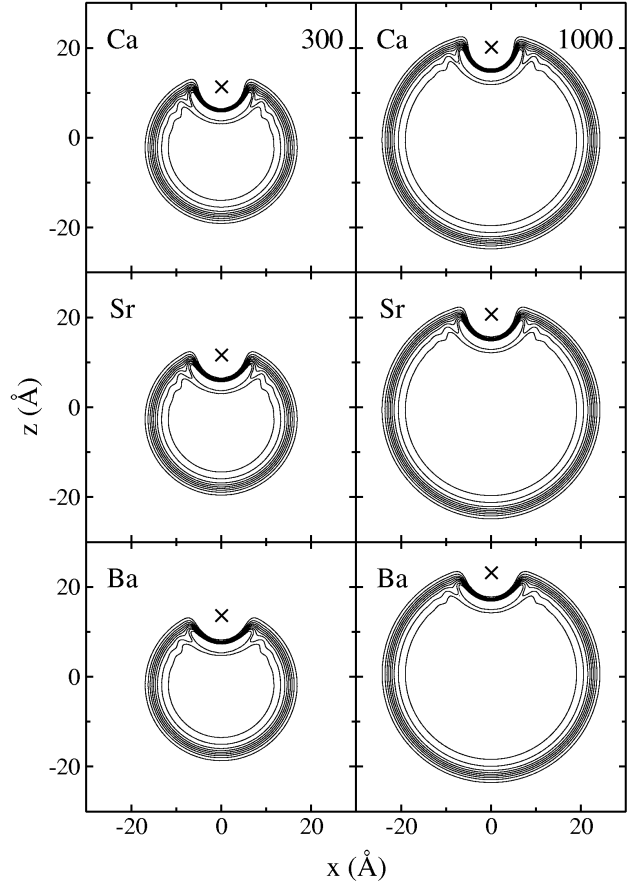}}
\caption{Equidensity lines on a symmetry plane for $^4$He$_N$ drops
with $N=$300 (left panels) and 1000 (right panels) doped with Ca,
Sr and Ba. The lines span the surface region between 0.9$\rho_b$ and
0.1$\rho_b$ in 0.1$\rho_b$ steps, where $\rho_b$ is the bulk liquid
density 0.0218~\AA$^{-3}$. The cross indicates the location of the
alkaline earth atom in the dimple. Results obtained using the pair
potentials of Ref.~\onlinecite{Lov04}.
}
\label{fig5}
\end{figure}

\begin{figure}[tb]
\centerline{\includegraphics[width=0.5\textwidth]{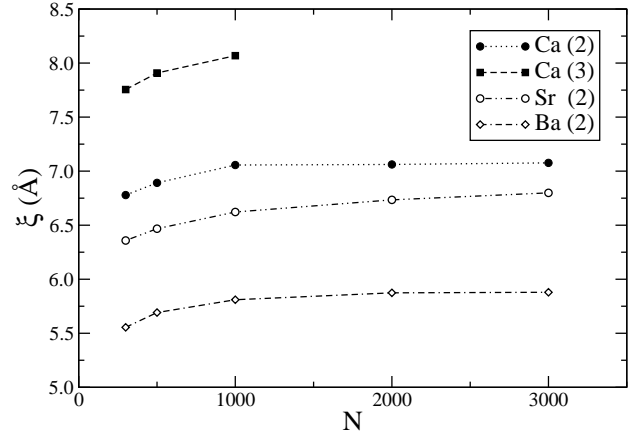}}
\caption{
Depth of the dimples ($\xi$) created in $^4$He$_N$ drops
obtained using the following pair potentials:
(2) from  Ref.~\onlinecite{Lov04} for Ba
(diamonds), Sr (circles), and Ca (solid dots) atoms;
(3) from  Ref.~\onlinecite{Meyer} for Ca (squares).
The lines are drawn to guide the eye.
}
\label{fig6}
\end{figure}

\begin{figure}[tb]
\centerline{\includegraphics[width=0.5\textwidth]{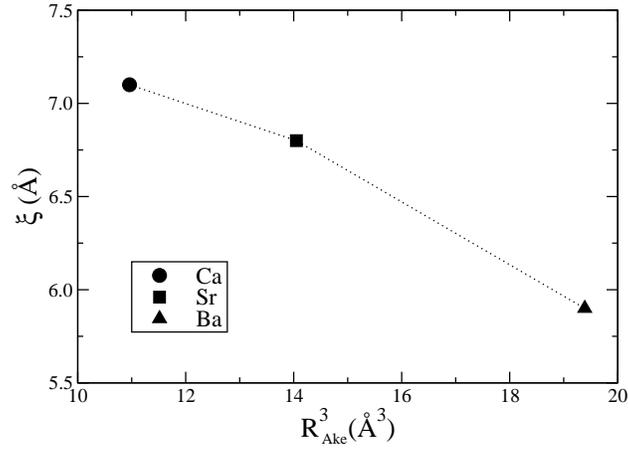}}
\caption{
Depth of the dimples ($\xi$) created in $^4$He$_{3000}$ drops by Ba,
Sr and Ca atoms, as a function of the atomic size $R^3_{Ake}$,
using the pair potentials of Ref.~\onlinecite{Lov04}.
The line is drawn to guide the eye.
}
\label{fig7}
\end{figure}

We have also confirmed by DF calculations the surface state of Ba@$^4$He$_N$ and the
solvated state of Mg@$^4$He$_N$, both suggested by the corresponding $\lambda$ values in
Table \ref{table1}. This is illustrated in Fig. \ref{fig4} for Mg, and in Fig. \ref{fig5}
for Ca, Sr and Ba. The dimple depth $\xi$, defined as the difference between the position
of the dividing surface at $\rho=\rho_b/2$ -where  $\rho_b$ is the bulk liquid density-
with and without impurity, respectively, is shown in Fig. \ref{fig6} as a function of
$N$. The structure of the dimple is
different for different alkaline earth atoms, being shallower for Ba and more pronounced
for Ca. We recall that the $\xi$ values for Na@$^3$He$_{2000}$ and Na@$^4$He$_{2000}$ are
4.5 and 2.1~\AA, respectively.\cite{Sti04} The dimple depths for alkaline are thus much
smaller than for alkaline earth atoms, as also indicated by LIF
experiments.\cite{Sti01,Sti04,Sti97,Sti99}
The dependence of the the dimple depth with the alkaline earth atom size,
characterized by the radial expectation value $R_{Ake}$ of the valence
electrons,\cite{Des73} is shown in Fig. \ref{fig7}. This figure is
consistent with the increasing bulk-to-surface ratio of the line
shifts as the size of the dopant atom increases.\cite{Sti99}

The `solvation' energies for these alkaline earth atoms in $^4$He drops are displayed in
the top panel of Fig. \ref{fig3}. As in the case of $^3$He drops discussed before, the
stronger the Ake-He pair potential (see Fig. \ref{fig1}), the more negative $S_N(Ake)$.
In the case of Ca@$^4$He$_N$ and Sr@$^4$He$_N$, the energies are very similar, and so are the dimple depths shown in Fig. \ref{fig6}. It is worth seeing the different behavior of
$S_N$ as a function of $N$ for each helium isotope. In the case of $^3$He, once the first
2-3 solvation shells are fully developed, $S_N$ quickly saturates, and for this reason
it changes only by 12 \% (Ca) and 17 \% (Sr) from $N=300$ to $N=5000$. For the same
reason, spectroscopic shifts are expected to be $N$ independent for drops made of more
than a few hundred $^3$He atoms. When the impurity is at the surface, sizeable curvature
effects appear even for a few thousand atoms drops. This shows up not only in the
change of $S_N$, which is about 22 \% for Ca, and 24 \% for Sr in the same $N$ range as
before, but also in the spectroscopic shifts, that still depend on $N$ below $N\sim
3000$ (see e.g. Ref.~\onlinecite{Sti99}). This illustrates the need of large drops
for carrying out spectroscopic shift calculations to attempt a detailed comparison with
experiments.

\section{Experimental results\label{iv}}

To support the DF calculations, the $5s5p~ ^1$P$_1^{\rm o} \leftarrow 5s^2~^1$S$_0$
transition of strontium on nanodroplets made of either helium isotope has been
experimentally investigated. Although calcium appears to be most favorable, we are so far
restricted to excitation spectra of strontium attached to helium droplets because of the
limited tuning range of our lasers. Calcium will be addressed in a future experiment. The
experiments where performed in a helium droplet machine applying laser-induced
fluorescence, as well as beam depletion and photo ionization (PI)
spectroscopy. A detailed description of the experimental setup is presented
elsewhere.\cite{Sti99} Modifications include a new droplet source to reach the lower
temperatures needed for generating $^3$He droplets.\cite{Sti04} In short, gas of either
helium isotope is expanded under supersonic conditions from a nozzle, forming a beam of
droplets traveling freely in high vacuum. The helium stagnation pressure in the droplet
source is $20$~bar, and a nozzle of $5$~$\mu$m diameter has been used. The
nozzle temperature has been stabilized to $12$~K and $15$~K to form $^3$He and $^4$He
droplets, respectively. These conditions result in an average droplet size of $\sim 5000$
helium atoms.\cite{Toe04}

The droplets are doped downstream
using the pick-up technique: in a heated scattering cell, an appropriate vapor pressure of
strontium is established so that droplets pick up one single atom on average when
passing the cell.
LIF as well as PI and BD absorption spectra of the doped droplet beam can be
recorded upon electronic excitation using a pulsed nanosecond dye laser. LIF is recorded
with a photo multiplier tube. In the case of PI, the photons of an excimer laser
ionize
the excited atoms in a one photon step. The ions are afterwards detected by a
channeltron. For the beam depletion measurement, a Langmuir-Taylor
surface ionization detector has been used.\cite{Sti00}

We show here only the results obtained using LIF because a much
better signal-to-noise ratio was achieved when compared to the PI
and BD spectra for strontium doped clusters. In the case of PI, the
reason probably is the tendency of the just formed strontium ions
not to desorb from the droplet like e.g. alkali atoms do. Since the
detection efficiency of our PI detector is considerably decreased
for high masses, the detection of the ion+droplet complex is small.
A decreased desorption mechanism also diminishes the sensitivity of
BD techniques. However, the PI/BD measurements give identical
results when compared to the LIF spectra.

\begin{figure}[tb]
\centerline{\includegraphics[width=0.5\textwidth]{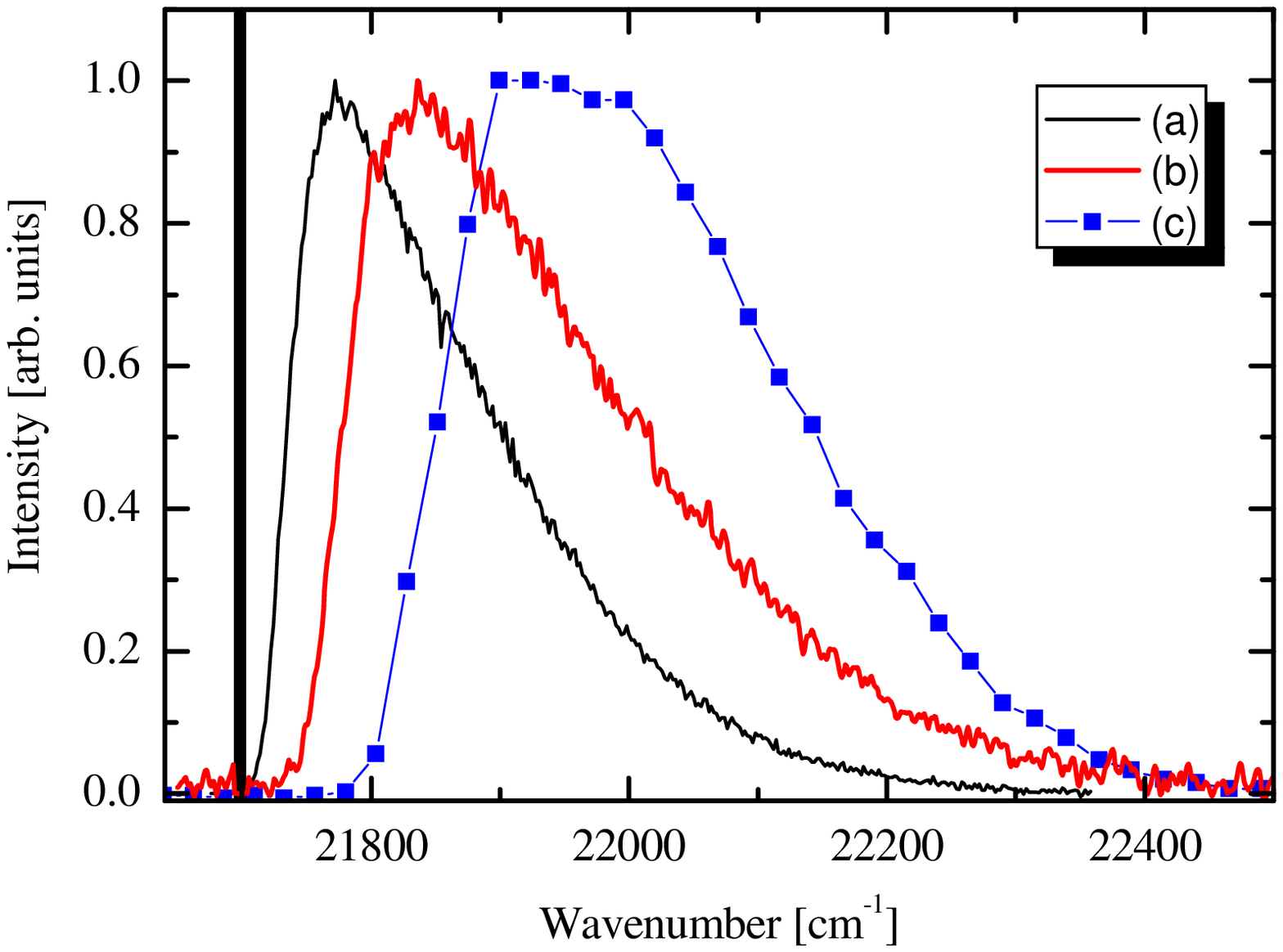}}
\caption{ Spectra of the Sr
$5s5p~ ^1$P$_1^{\rm o} \leftarrow 5s^2~^1$S$_0$ transition: (a) $^4$He drops, (b) $^3$He
drops, and (c) bulk $^4$He.\cite{Bau90} The vertical bar corresponds to
the atomic line.}
\label{fig8}
\end{figure}

Figure \ref{fig8} shows the measured spectra of the $5s5p~ ^1$P$_1^{\rm o} \leftarrow
5s^2~^1$S$_0$ transition of strontium atoms on droplets of $^3$He/$^4$He compared
to
that in bulk $^4$He.\cite{Bau90} All three spectra show a broad asymmetric line, blue
shifted from the atomic gas-phase absorption. The differences of the shifts for $^4$He
drops and bulk $^4$He immediately confirms the surface location of the
strontium
atoms.\cite{Sti97} In the case of bulk $^4$He, the absorption is far
more blue shifted and
the width is considerably wider. The shift can be explained within the
bubble model, see e.g. Refs. \onlinecite{Tab97,Mor05} and references
therein, and results from repulsion of the helium environment against spatial
enlargement of the electronic distribution of the excited state. The shift in bulk helium
is larger than in droplets because the dopant is completely surrounded by helium,
whereas it is not when it is located at the surface of drops.

Table \ref{table2} summarizes the experimentally determined shifts of the first
electronic transition of strontium and calcium in helium droplets, as well as the
measurements in bulk helium for both isotopes.
As compared to $^4$He drops, the absorption maximum in the case of $^3$He drops
is shifted $60$~cm$^{-1}$ further to the blue, and the
width increases from $180$ to $220$~cm$^{-1}$.

At first glance, it is not obvious from the recorded spectra in $^3$He drops
whether the strontium atom is in a surface state, or it is solvated inside the
droplets.  It is worth mentioning that Morowaki et
al. performed similar measurements in bulk helium.\cite{Mor05}
They have compared the
absorption spectra of the $4s4p~ ^1$P$_1^{\rm o} \leftarrow 4s^2~^1$S$_0$ transition of
calcium in bulk $^3$He and $^4$He, and have found a much smaller blue shift in the case of
$^3$He (about 55\%, see Table \ref{table2}),
which could again be explained within the bubble model -the
reduced shift just results from the lower density of liquid $^3$He.
A similar quantitative effect should be expected for strontium, especially in view of the
reported DF calculations.

Consistent with this expectation is that in our experiments,
the measured shift of Sr in $^3$He droplets, $140$~cm$^{-1}$, is about
a 58\% of the value corresponding to Sr in bulk
$^4$He, $240$~cm$^{-1}$.\cite{Bau90} We can safely argue that the
shift determined in $^3$He droplets should sensibly coincide with the expected
value for bulk $^3$He, indicating complete solvation
of strontium atoms in $^3$He droplets, as predicted by DF calculations.

\section{Summary and outlook\label{v}}

In this work, we have presented detailed results for the
structure and energetics of helium drops doped with Mg, Ca,
Sr, and Ba alkaline earth atoms. We have found that these atoms
are solvated in the case of $^3$He drops and reside in surface
dimples in the case of $^4$He drops, with the sole exception of
Mg@$^4$He$_N$, which is also solvated. This yields a fairly complete
physical picture, from the theoretical viewpoint, of the structure
and energetics of helium drops doped with alkaline earth atoms.
The experimental spectrum of strontium atoms in $^4$He and $^3$He
droplets confirms the DF calculations. Moreover, since the
spectroscopic shift is sensitive to the shape/depth of the surface
dimple, a comparison between experimental and calculated line
shifts could provide a sensible test on the accuracy of available
pair potentials. We want to stress again that accurate pair
potentials are needed to quantitatively reproduce the experimental
results, especially when the solvation properties of the
impurity are such that they yield values of $\lambda$ close to the
threshold value $\lambda_0$.

The different solvation behavior of the heavier alkaline earth atoms
in $^3$He and $^4$He drops, offers the unique possibility of using
them to study mixed drops at very low temperatures, in particular
the $^3$He-$^4$He interface. It is known that below the tricritical
point at $\sim0.87$~K,\cite{Edw92} $^3$He has a limited solubility
in $^4$He, segregating for concentrations larger than a critical
value. This segregation also appears in mixed
droplets,\cite{Gre98,Pi99,Fan05} producing a shell structure in
which a core, essentially made of $^4$He atoms, is coated by $^3$He
that is hardly dissolved into the $^4$He core, even when the number
of $^3$He atoms is very large.\cite{Pi99} Due to this particular
structure, that pertains to medium to large size droplets, strongly
attractive impurities reside in the $^4$He core, being very little
affected by the outer $^3$He shell, whereas weakly attractive
impurities, like alkali atoms, should still reside in the surface of
the droplet, irrespective of the existence of the $^4$He core.
Contrarily, Ca, Sr and Ba impurities would be sunk into the
fermionic component up to reaching the $^3$He-$^4$He interface if
the appropriate number of atoms of each isotope is chosen. This will
offer the possibility of studying the $^3$He-$^4$He interface, and a
richer alkaline earth atom environment. We are at present
generalizing the DF approach we have used in the past\cite{Pi99} to
address this more demanding and promising new aspect of the physics
of doped helium droplets. On the experimental side, calcium
spectra will be accessible in forthcoming experiments.
We want to point out that mixed droplets doped with alkali atoms
have been already detected in our previous experiments,\cite{Bue06}
and that systematic experiments on alkaline earth doped mixed droplets
will be performed in the future.

\section*{Acknowledgments}

We would like to thank Josef Tiggesb\"aumker and Marek Kro\'snicki for
useful correspondance. This work has been
performed under Grant No. FIS2005-01414 from DGI, Spain (FEDER), Grant 2005SGR00343 from
Generalitat de Catalunya, and under the HPC-EUROPA project (RII3-CT-2003-506079), with
the support of the European Community - Research Infrastructure Action under the FP6
`Structuring the European Research Area' Programme.

\newpage

\begin{table}[tb]
\caption{
$\lambda$ parameter for the alkaline earth atoms and pair potentials used
in this work.
}
\begin{center}
\begin{tabular}{lcc}
\hline
&\multicolumn{2}{c}{$\lambda$}\\
\cline{2-3}
&~~~$^3$He~~~ &~~~$^4$He~~~ \\
\hline
~~~${\mathrm{Mg}}^{a}$ & $4.73$ & $2.60$\\
~~~${\mathrm{Ca}}^{a}$ & $3.78$ & $2.08$\\
~~~${\mathrm{Ca}}^{b}$ & $3.71$ & $2.04$\\
~~~${\mathrm{Ca}}^{c}$ & $4.02$ & $2.21$\\
~~~${\mathrm{Ca}}^{d}$ & $4.52$ & $2.49$\\
~~~${\mathrm{Sr}}^{b}$ & $3.48$ & $1.92$\\
~~~${\mathrm{Ba}}^{b}$ & $3.15$ & $1.73$\\
\hline
\end{tabular}
\end{center}
\label{table1}
$^a$~Ref.~\onlinecite{Hin03}. $^b$~Ref.~\onlinecite{Lov04}.
$^c$~Ref.~\onlinecite{Meyer}. $^d$~Ref.~\onlinecite{Czu03}.
\end{table}

\begin{table}[tb]
\caption{Experimental shifts of the first electronic transition of
Ca and Sr atoms in bulk helium as well as in drops. The values for
Sr@He$_N$ are from this work. Previous experiments, carried out
only for Sr@$^4$He$_N$, showed the same shifts.\cite{Sti99}}
\begin{center}
\begin{tabular}{lcccc}
\hline
& \multicolumn{2}{c}{bulk} & \multicolumn{2}{c}{drop} \\
\cline{2-3}\cline{4-5}
& ~~~$^4$He~~~ & ~~~$^3$He~~~ & ~~~$^4$He~~~ & ~~~$^3$He~~~\\
\hline
${\mathrm{Ca}}$&&&&\\
\hfill shift(cm$^{-1}$)~~~& 203$^{b}$ & 112$^{b}$ &  72$^{a}$ & $-$ \\
\hfill FWHM(cm$^{-1}$)~~~& 297$^{b}$ & 245$^{b}$ & 173$^{a}$ & $-$ \\
\hline
${\mathrm{Sr}}$&&&&\\
\hfill shift(cm$^{-1}$)~~~& 240$^{c}$ & $-$       &  80       & $140$ \\
\hfill FWHM(cm$^{-1}$)~~~& 287$^{c}$ & $-$       & 180       & $220$ \\
\hline
\end{tabular}
\end{center}
\label{table2}
$^a$~Ref.~\onlinecite{Sti99}. $^b$~Ref.~\onlinecite{Mor05}. $^c$~Ref.~\onlinecite{Bau90}.
\end{table}

\newpage
\newpage
\newpage
\newpage
\newpage
\newpage
\pagebreak
\newpage


\begin{thebibliography}{99}
\bibitem{note}
Since we treat the impurity as an infinitely heavy particle, what actually
occurs is that during the functional minimization, helium is drawn towards the
impurity, embedding it until a final, lowest energy configuration is reached
where the impurity sits in the center of the droplet.
Numerically, the process is optimized by computing the force acting on
the impurity, and moving it in the direction of that force, see e.g.
Ref.~\onlinecite{Anc95}.
\end{thebibliography}
\end{document}